\documentclass{aa}
\usepackage{graphicx}
\usepackage{appendix}
\usepackage{txfonts}
\usepackage{auto-pst-pdf}

\begin{document}

\title{Interstellar ice analogs: band strengths of H$_2$O, CO$_2$, CH$_3$OH, and NH$_3$ in the far-infrared region.} 
\author{ B. M. Giuliano \inst{1}, R. M. Escribano \inst{2}, R. Mart\'in-Dom\'enech \inst{1}, E. Dartois \inst{3,4}, G. M. Mu\~{n}oz Caro \inst{1}}

\offprints{Barbara M. Giuliano}
\institute{Centro de Astrobiolog\'{\i}a, INTA-CSIC, Carretera de Ajalvir,
km 4, Torrej\'on de Ardoz, 28850 Madrid, Spain\\
\and Instituto de Estructura de la Materia-Consejo Superior de Investigaciones Científicas (IEM-CSIC), 28006 Madrid, Spain\\
\and CNRS-INSU, Institut d'Astrophysique Spatiale, UMR 8617, 91405 Orsay, France
\and Universit\'e Paris Sud, Institut d'Astrophysique Spatiale, UMR 8617, b\^atiment 121, 91405 Orsay, France\\}
\date{Received - , 2014; Accepted - , 2014}
\authorrunning{B. M. Giuliano et al.}  
\titlerunning{FIR}

  \abstract
{Whereas observational astronomy now routinely extends to the far-infrared region of the spectrum, systematic laboratory studies are sparse. In particular, experiments on laboratory analogs performed through the years have provided information mainly about the band positions and shapes, while information about the band strengths are scarce and derivable principally from the optical constants.}
{We measure the band strengths in the far-infrared region of interstellar ice analogs of astrophysically relevant species, such as  H$_2$O, CO$_2$, CH$_3$OH, and NH$_3$, deposited at low temperature (8-10 $\mathrm{K}$), followed by warm-up, to induce amorphous-crystalline phase transitions when relevant.}
{The spectra of pure H$_2$O, NH$_3$, and CH$_3$OH ices have been measured in the near-, mid- and far-infrared spectroscopic regions using the Interstellar Astrochemistry Chamber (ISAC) ultra-high-vacuum setup. In addition, far-infrared spectra of NH$_3$ and CO$_2$ were measured using a different set-up equipped with a bolometer detector. Band strengths in the far-infrared region were estimated using the corresponding near- and mid-infrared values as a reference. We also performed theoretical calculations of the amorphous and crystalline structures of these molecules using solid state computational programs at density functional theory (DFT) level. Vibrational assignment and mode intensities for these ices were predicted.} 
{Infrared band strengths in the 25-500 ${\mu}$m range have been determined for the considered ice samples by direct comparison in the near- and mid-infrared regions. Our values were compared to those we calculated from the literature complex index of refraction. We found differences of a factor of two between the two sets of values.}
{The calculated far-infrared band strengths provide a benchmark for interpreting the observational data from future space telescope missions, allowing the estimation of the ice column densities.}


\maketitle

%
%
\section{Introduction}
\label{intro}
\indent\indent 

Water and other simple molecules, such as methanol, carbon dioxide, carbon monoxide, methane, and ammonia, which are present in the interstellar medium (ISM) in the gas phase, can condense on the surface of dust grains under particular conditions of temperature and density, or can be directly formed by reactions. These processes usually take place in dense molecular clouds and circumstellar envelopes.

The molecular composition of the condensed phase can be investigated using infrared (IR) spectroscopy, by comparing the data from the observations with laboratory reference spectra. The near- and mid-infrared regions (extending typically from 1 ${\mu}$m to 25 ${\mu}$m) have been mainly used for the detection of molecular components in icy grain mantles. These regions provide information about the vibrational modes of the molecules, which can be used to identify and characterize the carriers of the IR bands.

The far-IR spectroscopic region (25-500 $\mu$m) is commonly associated with the intermolecular vibrational modes and can be used to unravel the structure and the transition between phases of a condensed medium. Astronomical observations in this region are limited (ISO, see, e.g., Dartois et al. 1998 and Cohen et al. 1999; KAO, see, e.g., Omont et al. 1990; and Herschel/PACS), and the advent of future missions (SPICA) will provide a new set of data that will make this spectroscopic region increasingly accessible.

Laboratory studies are necessary to provide a benchmark for the analysis of the ISM ice profiles. Far-IR band positions and shapes are sensitive to the temperature history of the sample and radiative transfer. In particular, amorphous ice converts to the crystalline phase with increasing temperature. The spectra of amorphous and crystalline phases of a pure molecular ice have characteristic and unique profiles in the far-IR. In the mid-IR this is not always as prominent. For instance, the IR band of CO ice near 2139 cm$^{-1}$ does not change significantly when this amorphous solid becomes crystalline. One of the physical parameters that can be determined from the laboratory data is the amorphous-to-crystalline phase transition temperature. This information, together with information about the radiation environment, could permit the determination of the temperature constraints of ice mantles on ISM dust grains.

Spectroscopic studies of pure ice analogs for water and other astrophysically relevant species in the far-IR region were previously reported by Bertie \& Whalley (1967), Brown \& King (1970), Ferraro et al. (1980), Sill et al. (1980), Warren (1984, and 1986), Martonchik et al. (1984), Anderson et al. (1988), Moore \& Hudson (1992), Hudgins et al. (1993), Smith et al. (1994), Johnson \& Atreya (1996), Trotta (1996), Schmitt et al. (1998), Maldoni et al. (1999), Coustenis et al. (1999), and Moore et al. (2001). These studies focused mainly on the determination of optical constants and the analysis of the spectral differences between the amorphous and crystalline structures. In addition, far-IR spectra of mixed ices of astrophysically relevant species have been investigated by Moore \& Hudson (1994). 

In the present work we have recorded the infrared spectra simultaneously in the near-, mid-, and far-IR regions of pure amorphous and crystalline  water, carbon dioxide, ammonia, and methanol ices. Carbon monoxide, and methane towards some lines of sight, are also abundant components in inter- and circumstellar ice mantles, but these species do not display absorption bands in the region above 100 cm$^{-1}$ (see Moore \& Hudson 1994). The main purpose of our analysis is to calculate the IR band strengths in the 25-500 ${\mu}$m spectral region for the considered species. For this, we performed spectroscopy of the same ice sample in the near- to far-IR range in order to estimate the band strengths in the far-IR based on their known values in the near- and mid-IR. This could allow the determination of the absorbing column densities in far-IR observations of ice mantles.

\section{Methods}\label{expe}

\indent\indent Ice analogs were prepared from gas condensation onto a window cooled at 8 $\mathrm{K}$, placed in the ultra-high-vacuum (UHV) compartment of the Interstellar Astrochemistry Chamber (ISAC) set-up described in  Mu\~{n}oz Caro et al. (2010).
The whole spectrum in the 6000-50 cm$^{-1}$ spectroscopic region was recorded in the same experiment, after special modifications to the ISAC set-up. In particular, the UHV chamber was equipped with two external and one internal diamond windows, which are transparent to the IR radiation in the whole spectroscopic range, except a small two-phonon related absorption in the 2000-2500 cm$^{-1}$ region. The absorption in this region, however, is compensated by subtracting the window spectrum recorded before the ice deposition.

Ice samples were formed by condensation onto the substrate of gaseous samples. The ice samples were subsequently heated in a controlled way, at a rate of 5 K per minute to the temperature at which the ices showed spectral changes. The evolution of the ice samples was monitored by Fourier transform infrared (FTIR) transmittance spectroscopy using a Vertex 70 spectrometer equipped with a deuterated triglycine sulfate detector (DTGS). The detector is placed outside the spectrometer at a distance of approximately 20 cm from the center of the substrate diamond window situated inside the chamber. This configuration was chosen because of the large dimension of the UHV chamber which cannot be allocated in the sample compartment of the spectrometer. In the near- and mid-IR regions all the spectra were recorded with a resolution of 2 cm$^{-1}$, aperture 6 mm, and averaged over 128 scan accumulations while in the far-IR region the spectra were recorded with a 4 cm$^{-1}$ resolution and 512 accumulations. The baseline was corrected manually using a spline function.

For the experiments carried out in Madrid, H$_2$O (triply distilled), NH$_3$ (supplied by Praxair, 99.999\% pure), and CH$_3$OH (supplied by Panreac, HPLC-gradient grade) were used without further purification.

Additional NH$_3$ and CO$_2$ experiments were performed at the Institut d'Astrophysique Spatiale (IAS). Mid- and far-IR spectra were recorded with an IFS66V Bruker Fourier transform spectrometer, coupled to a high-vacuum vessel (P $\ll$ 10$^{-7}$ mbar). The external port of the detector chamber was equipped with a 42 $\mathrm{K}$ IRLabs bolometer (for experiments in the far-IR region) or a DTGS CsI detector (in the mid-IR region). In the case of NH$_3$ ice, experiments were performed by deposition of ammonia vapor onto a cold CsI window. Spectra were recorded at a resolution of 1 cm$^{-1}$, aperture 6mm, and averaged over 500 scans. The CO$_2$ ice spectrum in the 200-50 cm$^{-1}$ region was recorded using a bolometer because the DTGS noise level in the frequency region where CO$_2$ absorbs was too high; these spectra were recorded at a resolution of 0.5 cm$^{-1}$, with an aperture of 8mm, and averaged over 1024 scans. High-frequency fringes appeared in the spectra and were filtered out. A Lee filter (see Lee, 1986) with a width of five spectral elements was applied, filtering out the Fabry-Perot effect variations between uncoated and coated substrate, due to the finite length of the substrate and higher spectral resolution for the recording of these spectra (0.5 to 1 cm$^{-1}$). The baseline was extrapolated from various thickness depositions. The individual spectra in both configurations overlap in the 240-700 cm$^{-1}$ region.

For the work performed at IAS, NH$_3$ and CO$_2$ vapour (supplied by Air Liquide, 99.999\%  and 99.995\% pure, respectively) were deposited onto a cold finger on which was mounted a CsI or polyethylene (depending on the spectroscopic region) window cooled to 10 $\mathrm{K}$ by liquid helium transfer. Deposition rates in our experiments varied between approximately 1 ${\mu}$m h$^{-1}$ and 4 ${\mu}$m h$^{-1}$. Deposition rates in the high end of this range were used in the IAS experiments. These rates are in agreement with those of previous studies which we used as references for the band strengths in the near- and mid-infrared regions (see Table~\ref{band_str}).

The column density of the deposited ice was calculated using the classical formula\\
\begin{equation}
 N =\frac{1}{A}\int_{band}{\tau_{\nu}d\nu},
\label{column}
\end{equation}
where $N$ is the column density in cm$^{-2}$, $\tau_{\nu}$ the optical depth of the band, $d\nu$ the wavenumber differential in cm$^{-1}$, and $A$ the band strength in cm molecule$^{-1}$. 

The adopted band strengths used to normalize the entire spectrum are taken from Sandford \& Allamandola (1993) and Gerakines et al. (2005) for NH$_3$ ice in the mid- and near-IR regions, respectively; from Allamandola et al. (1992), Sandford \& Allamandola (1993), and Hudgins et al. (1993) for CH$_3$OH ice; from Gerakines et al. (2005) for H$_2$O ice; and Gerakines et al. (1995) for CO$_2$ ice. The used values are summarized in Table~\ref{band_str}. Other references for band strength values in the near- and mid-IR can be adopted to recalculate the band strengths in the far-IR using our data.

\begin{table*} [tb]
\centering
\caption{Band strength ($A$) values of measured bands in the near- and mid- IR regions used for determination of the $A$ values in the far-IR.}
\label{band_str}
\begin{tabular}{l c c c c}
\hline \hline
Ice & Frequency (cm$^{-1}$) & Mode & $A$ (cm molec$^{-1}$) & Reference \\ [0.5ex]
\hline
 &  &  &  & \\
{NH$_3$} & 4995 & ${\nu}_{\rm 1}$+${\nu}_{\rm 4}$ (?) & 8.1 $\times$ 10$^{-19}$ & a \\
 & 4480 & ${\nu}_{\rm 1}$+${\nu}_{\rm 2}$ (?) & 8.7 $\times$ 10$^{-19}$ & a \\
 & 3211-3370 & ${\nu}_{\rm 1}$,${\nu}_{\rm 3}$ & 2.2 $\times$ 10$^{-17}$ & b \\
 & 1625 & ${\nu}_{\rm 4}$ & 4.7 $\times$ 10$^{-18}$ & b \\
 & 1070 & ${\nu}_{\rm 2}$ & 1.7 $\times$ 10$^{-17}$ & b \\
 &  &  &  & \\
{CH$_3$OH} & 4396 & ${\nu}_{\rm 1}$+${\nu}_{\rm 11}$ or ${\nu}_{\rm 1}$+${\nu}_{\rm 7}$ & 8.7 $\times$ 10$^{-19}$ & a,b \\
 & 4271 & ${\nu}_{\rm 1}$+${\nu}_{\rm 8}$ & 8.0 $\times$ 10$^{-20}$ & a,b \\
 & 2527 & combination mode & 2.6 $\times$ 10$^{-18}$ & c \\
 & 1461 & ${\nu}_{\rm 4}$,${\nu}_{\rm 5}$,${\nu}_{\rm 6}$,${\nu}_{\rm 10}$ & 1.2 $\times$ 10$^{-17}$ & d \\
 & 1130 & ${\nu}_{\rm 7}$,${\nu}_{\rm 11}$ & 1.8 $\times$ 10$^{-18}$ & d \\
 & 705 & ${\nu}_{\rm 12}$ & 1.4 $\times$ 10$^{-17}$ & d \\
 &  &  &  & \\
{H$_2$O} & 5001 & ${\nu}_{\rm 2}$+${\nu}_{\rm 3}$ & 1.2 $\times$ 10$^{-18}$ & a \\
 &  &  &  & \\
{$^{13}$CO$_2$} & 2283 & ${\nu}_{\rm 3}$ & 7.8 $\times$ 10$^{-17}$ & e \\ [0.5ex]
\hline
\end{tabular}

\begin{list}{}
\item (a) From Gerakines et al. (2005).
\item (b) From Sandford \& Allamandola (1993). 
\item (c) From Allamandola et al. (1992).
\item (d) From Hudgins et al. (1993).
\item (e) From Gerakines et al. (1995).
\end{list}
\end{table*}

For each molecule the band strengths of the pure amorphous ice in the far-IR region were calculated as follows. The ice column density was estimated from the integrated optical depths of the bands in the near-  and mid-IR regions, using the corresponding known values of the strengths taken from the literature. Once the ice column density was estimated, the band strengths in the far-IR region were derived using the equation\\
\begin{equation}
 {A_{FIR}}=\frac{1}{N}\int_{band}{\tau_{\nu}d\nu}=\frac{\int_{band}^{FIR}{\tau_{\nu}d\nu}}{\int_{band}^{MIR}{\tau_{\nu}d\nu}}A_{MIR}
\label{bandstr}
\end{equation}
which follows from Eq.~\ref{column}.

 For the ices in their crystalline phases, the band strengths were calculated from the corresponding values in the amorphous phase by scaling the relative integrated optical depth.

 We also carried out a number of theoretical calculations to help in the analysis of the laboratory data. The calculations were done using the CASTEP (Clark et al. 2005) and Amorphous Cell packages of Materials Studio (MS) (http://accelrys.com/products/materials-studio/, Segall et al. 2002) for all samples, except for amorphous water, for which we employed the SIESTA method (Soler et al. 2002, Ordej\'{o}n et al. 1996, and Gonze \& Lee 1997) to make use of the more adequate Van der Waals-type functionals (Dion et al. 2004). The calculations performed in this work involved the relaxation of initial geometries until a minimum on the potential energy surface was found, for which a vibrational analysis including the prediction of the vibrational spectrum was carried out. Details on the specific method and parameters used for each calculation are given in Table~\ref{table_calc}. In the present work, calculations were carried out for both the crystalline and amorphous phases of the ices studied experimentally. The initial geometry for the crystalline structures was the unit cell of each species taken from crystallographic or neutron scattering experiments, while for the amorphous structures initial geometries were prepared using the MS Amorphous Cell module, fixing the density of the sample to the best available experimental value.

\begin{table*} [tb]
\centering
\caption{Details of the calculations carried out for the ices studied in this work.}
\label{table_calc}
\begin{tabular}{l c c c c c}
\hline \hline
Method & Functional, basis & Energy tolerance & Stress tolerance & Force tolerance & Atom displacement tolerance \\ [0.5ex]
 &  & (eV/atom) & (GPa) & (eV/A) & (A) \\
\hline
{CASTEP} & GGA PBE$^a$, plw$^b$ & 5.0 $\times$ 10$^{-6}$ & 0.02 & 0.01 & 0.0005 \\ [0.5ex]
{SIESTA} & VDW$^c$ , opt$^d$  & 4.2 $\times$ 10$^{-6}$ & 0.1 & 0.0005 & 0.05 \\ [0.5ex]
\hline
\end{tabular}

\begin{list}{}
\item $^a$ GGA stands for Generalized Gradient Approximation, and PBE for Perdew-Burke-Ernzerhof functionals, see Perdew et al. (1996).
\item $^b$ CASTEP uses plane wave basis sets.
\item $^c$ Van der Waals adapted functionals, see Dion et al. (2004).
\item $^d$ Specifically optimized basis set, see Fern\'andez-Serra \& Artacho (2004).
\end{list}
\end{table*}

\section{Experimental results}
\label{results} 
\indent\indent The experimental results of deposition and subsequent heating of pure ices are described below. The ices were deposited at low temperature as amorphous solids. In this disordered phase, molecules are not oriented in the most energetically favorable position, unless sufficient energy is provided to the system. Annealing of the ice promotes the molecules to a higher energy state, allowing them to rearrange into more ordered phases, which are energetically advantageous. This phase transition produces changes in the infrared spectra that are indicative of the physical state of the monitored solid. In the present experiments, the deposited ices were heated up to the phase transition temperature in order to produce their crystalline phases.

 In the next sections, the procedure adopted for each molecular species is described in more details. The derived band strengths and tentative identifications of the corresponding vibrational modes are given in Table~\ref{table_all}.

\begin{table*} [tb]
\centering
\caption{Infrared band positions and band strength ($A$) values of monocomponent ices in the range of interest from 500 cm$^{-1}$ to 50 cm$^{-1}$ at the deposition and phase transition temperatures.}
\label{table_all}
\begin{tabular}{l c c c c l l}
\hline \hline
Ice & Temperature ($\mathrm{K}$) & Frequency (cm$^{-1}$)/(${\mu}$m) & Mode & Mode$^a$ & $A$ (cm molec$^{-1}$) & $A$ (cm molec$^{-1}$)$^b$ \\ [0.5ex]
\hline
 &  &  &  &  &  & \\
{NH$_3$} &  &  &  &  &  & \\
amorphous & 10 & 419 (22.2) & - & - & 8.1 (1.5)$\times$ 10$^{-18}$ & 1.4 $\times$ 10$^{-17}$ \\
crystalline & 100 & 535 (19.1) & - & wagging &  & \\ [-1ex]
 &  & 421 (27.5) & librational $^c$ & torsion & \raisebox{1.5ex}{1.2 (0.2)$\times$ 10$^{-17}$ $^{d,e}$} & \raisebox{1.5ex}{2.0 $\times$ 10$^{-17}$} \\ [1ex]
 &  &  &  &  &  & \\
{CH$_3$OH} &  &  &  &  &  & \\
amorphous & 8 & 303 (33.0) & - & - & 9.3 (4.8)$\times$ 10$^{-20}$ & 2.0 $\times$ 10$^{-19}$ \\
crystalline & 120 & 347 (28.8) & translational $^f$ & librational & 1.3 (6.6)$\times$ 10$^{-18}$ $^d$ & 5.1 $\times$ 10$^{-19}$ \\
 &  & 177 (56.5) & librational $^f$ & translational & 8.2 (4.3)$\times$ 10$^{-19}$ $^d$ & 3.9 $\times$ 10$^{-19}$ \\ [0.5ex]
 &  &  &  &  &  & \\
{H$_2$O} &  &  &  &  &  & \\
amorphous & 8 & 219 (45.7) & - & - & 4.4 (1.7)$\times$ 10$^{-18}$ & 4.9 $\times$ 10$^{-18}$ \\
crystalline & 150 & 227 (44.1) & transverse optical $^g$ & librational & \\ [-1ex]
 &  & 160 (62.5) & longitudinal acoustic $^g$ & translational & \raisebox{1.5ex}{5.3 (2.0)$\times$ 10$^{-18}$ $^{d,e}$} & \raisebox{1.5ex}{5.5 $\times$ 10$^{-18}$} \\ [1ex]
 &  &  &  &  &  & \\
{CO$_2$} &  &  &  &  &  & \\
amorphous & 10 & 117 (85.5) & translational $^h$ & translational & 1.3 (0.2)$\times$ 10$^{-19}$ & - \\
 &  & 69 (144.9) & - & translational & 3.5 (0.6)$\times$ 10$^{-20}$ & - \\ [0.5ex]
\hline
\end{tabular}

\begin{list}{}
\item $^a$ Description from theoretical calculations (see text).
\item $^b$ Derived from the optical constants available in the literature (see Section~\ref{disc}). 
\item $^c$ From Ferraro et al. (1980).
\item $^d$ Estimated errors in the band strengths of the crystalline forms are derived from the corresponding percentual error calculated for the amorphous ice.
\item $^e$ The band strength has been calculated for the whole band including the two components.
\item $^f$ From Anderson et al. (1988).
\item $^g$ From Bertie \& Whalley (1967).
\item $^h$ From Brown \& King (1970).
\end{list}
\end{table*}


 With the calculations, we also aimed to achieve a graphical picture of the vibrational modes, especially the lattice and low-frequency modes, and to relate them to their corresponding bands in the spectra, particularly in the far-IR region. In most cases, we focused on the calculations of the crystalline species, which are easier to visualize. In addition, the vibrations for the amorphous structures followed in general the same trend as for the corresponding crystals, but without the phase characteristics typical of crystals. Instead, individual molecules vibrate on their own within a given normal mode, yielding slightly different frequencies that broaden the spectral band. 

 A word of caution must be added regarding the comparison between observed and calculated spectra. Most solid state calculations are based on the harmonic approximation models for the geometry and force field of the system. This implies that the calculations provide the frequency and intensity of the normal modes, but no information is available with these models on the width of the corresponding infrared bands, an essential property that configures the spectral appearance. Because this information is missing, all theoretical spectra were broadened with a Lorentzian function of half-width at half-maximum of 20 cm$^{-1}$, whereas each calculated mode ought to be broadened by its appropriate width function to allow a proper comparison. Apart from this, the predicted frequencies match the observed ones fairly well.

 Comments on each case are given at the end of the appropriate sections.


\subsection{NH$_3$ ice}
\label{results_NH3}
\begin{figure}  [ht!]
   \centering
    \includegraphics[width=10.cm]{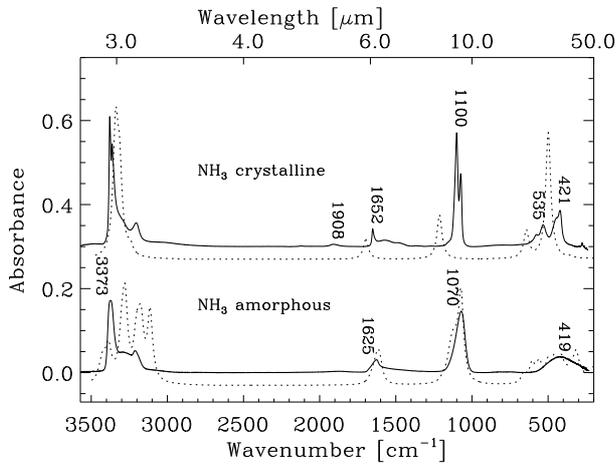} 
     \caption{Experimental (solid line) and calculated (dotted line) IR spectrum of NH$_3$ ice. The laboratory spectra of amorphous NH$_3$ was deposited and recorded at 10~$\mathrm{K}$, while the crystalline spectrum of NH$_3$ was obtained after warming up to 100~$\mathrm{K}$.}
      \label{NH3}
\end{figure}

\indent\indent The spectrum recorded at IAS of solid NH$_3$, deposited at 10~$\mathrm{K}$ and warmed up to 100~$\mathrm{K}$, is shown in Fig.~\ref{NH3}.

 The shape of the far-IR bands of amorphous and crystalline NH$_3$ ice is analogous to previous results shown by Moore \& Hudson (1994). We performed three different experiments with different ice thickness varying from approximately 1 $\mu$m to 6 $\mu$m. The ice thickness was estimated from the calculated column densities (varying from 2.1 $\times$ 10$^{18}$ molecules cm$^{-2}$ to 1.3 $\times$ 10$^{19}$ molecules cm$^{-2}$) and assuming a density of 0.66 g cm$^{-3}$ (taken from Satorre et al. 2013). The column densities were calculated using the integrated optical depth of the bands in the near- and mid-IR at 1070, 1625, 3373, 4480, and 4995 cm$^{-1}$. Then, the 419 cm$^{-1}$ band strength value was calculated as arithmetic mean of the three different experiments. As we can see from the associated error, the three values are very consistent. The strength of the bands of the crystalline form were derived from the value of the amorphous phase, by scaling the relative integrated optical depth. The associated error was estimated using the calculated error for the band strength of the amorphous phase.

 The calculated spectrum of the crystal seems to be shifted with respect to the observation by $\sim$ 100 cm$^{-1}$ in the low-frequency region, whereas the calculated spectrum of the amorphous solid agrees well with the observation. Both calculations are carried out using a similar method and parameters, which makes it hard to put forward an explanation for this shift. The strong vibration measured at 421 cm$^{-1}$ (calculated at 500 cm$^{-1}$), described by Ferraro et al. (1980) as librational, should correspond to a kind of internal rotation or torsion of two of the N-H bonds with respect to the third one, and the band observed at 535 cm$^{-1}$ (calculated at 643 cm$^{-1}$) can be described as an in-phase internal rotation or wagging of all three N-H bonds. In the amorphous structure, these vibrations are mixed, resulting in a broad feature, both in the observation and in the prediction, where the calculated contour is somewhat broader than the observed one.

\subsection{CH$_3$OH ice}
\label{results_CH3OH}

\begin{figure}  [ht!]
   \centering
    \includegraphics[width=10.cm]{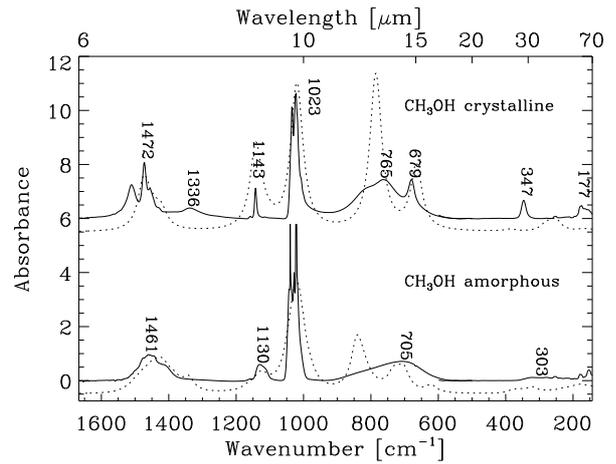} 
     \caption{Experimental (solid line) and calculated (dotted line) IR spectrum of CH$_3$OH ice. The laboratory spectra of amorphous CH$_3$OH was deposited and recorded at 8~$\mathrm{K}$, while the crystalline spectrum of CH$_3$OH was obtained after warming up to 120~$\mathrm{K}$.}
      \label{CH3OH}
\end{figure}

\indent\indent Figure~\ref{CH3OH} shows the IR spectrum of CH$_3$OH ice, deposited at 8~$\mathrm{K}$ and warmed up to 120~$\mathrm{K}$. Table~\ref{table_all} shows the peak position and assignment of the far-IR components. The ice thickness has been estimated to be approximately 13 $\mu$m from the calculated column density (2.5 $\times$ 10$^{19}$ molecules cm$^{-2}$), assuming an ice density of 1.013 g cm$^{-3}$  estimated for 80~$\mathrm{K}$ (from Mat\'e et al. 2009). The far-IR bands profile is consistent with the one shown by Moore \& Hudson (1994) and Hudgins et al. (1993) for both amorphous and crystalline phases.

 In the far-IR region the strength of the 303 cm$^{-1}$ band was determined from the integrated optical depth of the bands in the near- and mid-IR regions, at 705, 1130, 1461, 2527, 4271, and 4396 cm$^{-1}$, whose band strength values are given in Table~\ref{band_str}. The strength of the bands of the crystalline form, at 177 cm$^{-1}$ and 347 cm$^{-1}$, were derived from the value of the strength of the 303 cm$^{-1}$ band, by scaling the relative integrated optical depth. The associated errors for all values were estimated from the error associated to the calculation of the ice column density.

 As it is for NH$_3$, the calculated spectrum of the amorphous is in better agreement with the experiment than that of the crystal in the region $<$ 500 cm$^{-1}$. The band at 347 cm$^{-1}$ in the crystal, much broadened in the amorphous, could be assigned to a librational mode according to our calculations for the amorphous structure, rather than as a translational mode as advanced by Anderson et al. (1988). Similarly, our results for the amorphous describe the lower frequency peaks at $\sim$ 175 cm$^{-1}$, which are also of a librational nature. We only see a certain kind of translational motion in very weak features of the crystal at 170 cm$^{-1}$, unnoticeable in the compressed graph of Fig.~\ref{CH3OH}.

\subsection{H$_2$O ice}
\label{results_H2O}

\begin{figure}  [ht!]
   \centering
    \includegraphics[width=10.cm]{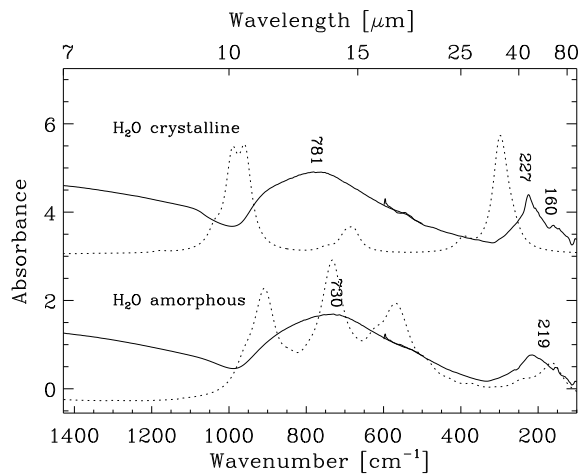} 
     \caption{Experimental (solid line) and calculated (dotted line) IR spectrum of H$_2$O ice. The laboratory spectra of amorphous H$_2$O was deposited and recorded at 8~$\mathrm{K}$, while the crystalline spectrum of H$_2$O was obtained after warming up to 150~$\mathrm{K}$.}
      \label{H2O}
\end{figure}

\indent\indent  The IR spectrum of a pure H$_2$O ice, deposited at 8~$\mathrm{K}$ and warmed up to 150~$\mathrm{K}$ is shown in Fig.~\ref{H2O}. The peak position and assignment of the far-IR components are shown in Table~\ref{table_all}. The spectroscopic pattern in the far-IR region of amorphous and crystalline phases is in agreement with earlier publications by Moore \& Hudson (1992) and Smith et al. (1994). The intensity of the far-IR bands is considerably smaller than of the ones in the mid-IR region. Figure~\ref{H2O} shows that even the intrinsically weak libration band around 760 cm$^{-1}$ in the mid-IR is stronger than the far-IR bands at frequencies below 300 cm$^{-1}$. Therefore, in order to obtain a significant signal-to-noise ratio in these regions, the bands in the mid-IR region become extremely saturated and cannot be used for the band strength calculation.

 Three different experiments with ice thickness varying from 2 $\mu$m to 10 $\mu$m were performed. The ice thickness has been estimated from the integrated optical depth of the band in the near-IR region at 5040 cm$^{-1}$, assuming a density of 0.94 g cm$^{-3}$ for amorphous and crystalline H$_2$O ice (from Dohn\'{a}lek et al. 2003, and Raut et al. 2007). The crystalline form was analyzed only in the 10 $\mu$m ice because in thinner ices desorption was competing with phase transition as the temperature was increased. The strength of the 219 cm$^{-1}$ band and the associated error were determined as the arithmetic mean of the results from the three different experiments. The error thus gives an estimation of the dispersion of the three values. For these experiments the error in percentage is calculated to be around 37~\%. For the crystalline phase, the strength of the bands was derived from the value of the strength of the 219 cm$^{-1}$ band by scaling the relative integrated optical depth. The associated error was estimated using the calculated error for the amorphous phase.

 The predicted spectra for either cubic or hexagonal H$_2$O ice are fairly similar overall, but with some minor differences. In the low-frequency region the strongest mode appears at 301 cm$^{-1}$ in cubic ice, with a shoulder at 288 cm$^{-1}$, but it is a weaker feature at 292 cm$^{-1}$ in hexagonal ice, with a shoulder at 283 cm$^{-1}$. These modes are librational, with translation and rotational components. Furthermore, the prediction for cubic ice gives a very weak peak at 169 cm$^{-1}$ not visible in Fig.~\ref{H2O}; it is a pure translational mode, which has null intensity in hexagonal ice. The appearance of this peak in the observed spectra indicates that the crystallization process may not be complete at 150~$\mathrm{K}$.

\subsection{CO$_2$ ice}
\label{results_CO2}

\begin{figure}  [ht!]
   \centering
    \includegraphics[width=10.cm]{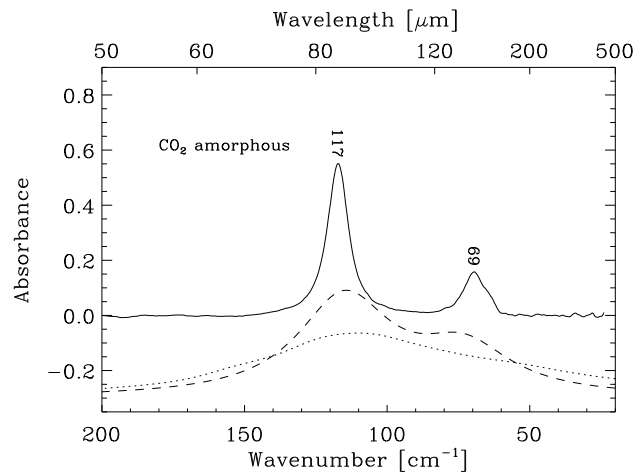} 
     \caption{Experimental IR spectrum (measured with a bolometer detector, solid line) and calculated IR spectrum (amorphous, dotted line; and crystalline, dashed line) of CO$_2$ ice. The laboratory spectra of amorphous CO$_2$ was deposited and recorded at 10~$\mathrm{K}$.}
      \label{CO2}
\end{figure}

\indent\indent The IR spectrum of a CO$_2$ pure ice, deposited at 10~$\mathrm{K}$, recorded at IAS using a bolometer detector, is shown in Fig.~\ref{CO2}. The peak position and assignment of the far-IR components are shown in Table~\ref{table_all}. One of our previous experiments measured the IR spectrum up to 700 cm$^{-1}$ and allowed us to investigate the integrated absorbance value for the 117 cm$^{-1}$ and 69 cm$^{-1}$ modes, with respect to the $^{13}$CO$_2$ bending mode, whose ratio is evaluated to be $\sim$ 1.05 $\times$ ($^{13}$C/$^{12}$C). The derived values are given in Table~\ref{table_all}.

Our calculations for amorphous CO$_2$ ice predict two peaks, one at ~ 125 cm$^{-1}$ and a much weaker one at ~ 66 cm$^{-1}$, in good agreement with the measured spectrum. They can both be described as translational modes. They are calculated at 115 cm$^{-1}$ and 74 cm$^{-1}$ for the pure crystal.

\section{Discussion}
\label{disc}

\indent\indent A detailed discussion of the profiles and shapes of the bands recorded in the far-IR region, as well as the physical properties of the amorphous and crystalline ice phases, which are temperature related, has been the subject of previous studies and is beyond the scope of this paper. The far-IR features are generally related to lattice modes, which involve the relative motion of the molecules as a whole. Therefore, these features are essentially related to intermolecular vibrations, either translational or librational modes, indicating respectively motions that can be described by displacements or torsions. These modes are phase dependent, hence, they can be used as a source of information on lattice structure and transition between phases. In this section we will summarize the spectroscopic behavior in the mid- and far-IR of each ice and how these features can be related to its structure. The effects of ice mixtures on these modes will be the subject of a dedicated manuscript which is in preparation.

In Moore \& Hudson (1994), the authors show that two of the most abundant ices, the CO and CH$_4$ ices do not display any feature in the far-IR, down to 100 cm$^{-1}$, and the features in the mid-IR are not always indicative of the physical state of the ice since, in some cases like pure CO ice, they do not show notable changes.
Ammonia ice displays band split, frequency shift, and change in the intrinsic intensity of the bands both in mid- and far-IR regions following state transition. In addition, new bands at approximately 1900 cm$^{-1}$ and 530 cm$^{-1}$ appear once the crystalline phase is formed.
Methanol ice samples show differences both in the mid- and far-IR. In the mid-IR, various bands split into two or more components when the physical state of the ice changes. In the far-IR, the changes of the band close to 300 cm$^{-1}$ are very pronounced both in terms of frequency shift and intrinsic intensity of the band.
The H$_2$O ice exhibits a profile change in the OH stretching band near 3200 cm$^{-1}$ when the ice phase changes from amorphous to crystalline, and a more pronounced change near 200 cm$^{-1}$, where the band splits into two components following ice phase transition. The transition from crystalline cubic to hexagonal H$_2$O ice is difficult to observe in the mid-IR, but, according to the results of the simulations presented in this paper, in the far-IR the presence or absence of the 160 cm$^{-1}$ band is an indication of cubic or hexagonal ice.
In the case of pure CO$_2$, only very thin ice samples display important changes in the mid-IR bands during the ice accretion or after warm-up (Escribano et al. 2013). In the far-IR region, the CO$_2$ ice spectrum does not change considerably; only narrowing of the band is observed in the transition from amorphous to crystalline. 

The prediction of simulated spectra for the far-IR region is complicated because the spectra depend strongly on the relaxed molecular arrangement in the ice, whereas, for the mid-IR, intramolecular vibrations dominate. A good match with the experimental data is therefore difficult to achieve. Nonetheless, we have been able to provide assignments for the observed bands.

To strengthen our analysis we also determined the absorption coefficients of the far-IR bands from the optical constants available in the literature. Optical constant data for water ices (at 33~$\mathrm{K}$ and 133~$\mathrm{K}$, respectively, for the amorphous and crystalline phases) and ammonia ices (at 15~$\mathrm{K}$ and 75~$\mathrm{K}$) have been downloaded using the GhoSST web interface (Schmitt et al. 2011), collected from Trotta's thesis (1996); while data for methanol ice (at 10~$\mathrm{K}$ and 120~$\mathrm{K}$) were collected from Hudgins et al. (1993). No data were found for the carbon dioxide ice in the 200-50 cm$^{-1}$ frequency region.

The corresponding band strengths were derived using the relation reported in Dartois \& d'Hendecourt (2001), \\

\begin{equation}
 {A}=\frac{1}{N}\int{\tau_{\nu}d\nu}=\int {4\pi k \nu d\nu} \frac{m}{\rho N_{A}},
\label{bandstrk}
\end{equation}
where $N$ is the column density, $k$ the imaginary part of the complex index of refraction, $\nu$ the wavenumber, $\rho$ the density of the material, $N_A$ the Avogadro constant, and $m$ the molecular weight of the constituent considered.

The integrated band strengths derived from Eq.~\ref{bandstrk} are an approximation that only considers the optical index $k$. The real part of the complex index of refraction ($n$) also affects the calculated absorbance spectrum (Hudson et al. 2014, and references therein). However, these differences between simulated and measured band strengths are expected to be minimized in thick ices (Hudson et al. 2014). The values obtained with Eq.~\ref{bandstrk} are thus comparable with those extracted from our laboratory experiments.

Using the density values listed in Section~\ref{results}, in the corresponding subsection for each ice, we obtained the values listed in Table~\ref{table_all}. As we can see, in most cases there is a fairly good agreement between the values derived from the optical constants and the values derived from the analysis of the bands in the mid- and near-IR regions presented in this paper. Differences on the order of a factor of two were typically found.

\section{Astrophysical implications}
\label{astro}

\indent\indent Laboratory data of ices in the IR provide a benchmark for the observations in various astrophysical environments. The optical depths of absorptions in the far-IR are predictable from the bands detected in the mid-IR for the same ice components, or alternatively, by using the band strength values measured in laboratory ice analogs. Then, the column densities of the molecules composing the ice, observed in dense interstellar clouds or circumstellar regions, can be determined from the band strengths.
Moreover, the strength of a certain absorption feature can vary depending on the physical phase of the ice, which can change from amorphous to crystalline and vice versa, depending on the thermal history of the sample and its radiation environment. 
On the basis of the results presented here, the determination of the column densities based on far-IR ice absorptions should be possible in various astrophysical objects.

So far, the only ice detected in the far-IR is water ice near 44 $\mu$m from ISO data (Omont et al. 1990, Dartois et al. 1998, and Cohen et al. 1999), but future planned missions like SPICA or Herschel/PACS archive data beyond 50 $\mu$m will expand the detection of ices to this wavelength range.

Lattice modes could, in principle, be observed in the spectra of segregated ice mantles, while the modes intrinsic to crystalline phases of pure ices will not be present in well-mixed ice mantles. An exception is the dominant H$_2$O ice component, since the overabundance of water molecules toward numerous lines of sight allows the crystallization of water ice. Indeed, the detection of the band at 44 $\mu$m, is an indication of crystalline water ice (Dartois et al. 1998).

So far, other molecular ice components detected routinely in the mid-IR have not been observed in far-IR spectra. This could be the result of mixing with other ice components, as mentioned above, but other effects can hinder the detection of weak absorption bands in the observed far-IR spectra. Ammonia should interact strongly when mixed with water, leading to structural and sometimes chemical changes like proton transfer, thus requiring the investigation of the far-IR spectrum of mixtures.

According to Dartois et al. (1998), temperature and opacity effects can be critical at wavelengths larger than 30 $\mu$m. According to these authors, the modification of the ice absorption band profiles due to emission can be significant in lines of sight toward protostellar environments, where an important temperature gradient is expected, while extinction can also induce important spectral alterations for these sources.

\section{Conclusions}
\label{conc}

\indent\indent In this article, we estimated the far-IR band strengths of astrophysically relevant molecules by comparison with the known near- and mid-IR features for the same ice sample. For this purpose, the ISAC set-up was equipped with special diamond windows to ensure transparency in the whole spectral range from 6000 cm$^{-1}$ to 50 cm$^{-1}$. Part of the spectra were recorded at IAS, which offered the advantage of a bolometer detector especially for the measurement of CO$_2$ ice. The absorption strengths were measured for the NH$_3$, CH$_3$OH, H$_2$O, and CO$_2$ pure ices, both in their amorphous phase at low temperature (8-10 $\mathrm{K}$) and crystalline phase at the corresponding phase transition temperature, which can vary depending on the ice composition. 

The analysis of the spectra is supported by theoretical calculations using solid state quantum chemical programs.

The imaginary part of the complex index of refraction for those species available in the literature was used to derive the band strengths in the far-IR, in order to present a comparison with our values, which are in fairly good agreement.

The presented data are of interest for observational astrophysics because the band strength values allow the column density calculation of pure molecular ice components. Our results, complemented with the study of ice mixtures, provide a benchmark for interpreting the data from future space telescope missions operating in the far-IR.

\begin{acknowledgements}
The authors acknowledge the referee, R. L. Hudson, for his useful comments. We acknowledge J. Cernicharo and M. Castellanos for their support on this project. We thank Bruker Spanish service for the technical assistance, in particular J. Bardaj\'{\i}, C. Villar and M. Crespo. B.~M.~G. acknowledges support from CONSOLIDER grant CSD2009-00038. R.~M.~D. benefited from a FPI grant from the Spanish MINECO. This research was financed by the Spanish MINECO under Projects AYA2011-29375 and CONSOLIDER grant CSD2009-00038 and FIS2010-16455.
\end{acknowledgements}


\end{document}